\documentclass[aps,pra,groupedaddress,showpacs,showkeys,twocolumn]{revtex4}

\usepackage{amsfonts}
\usepackage{amsmath}
\usepackage{amssymb}
\usepackage{graphicx}
\usepackage{epsfig}

\allowdisplaybreaks

\begin{document}


\title{Comment on ``Dynamics of open quantum systems initially entangled with environment: Beyond the Kraus representation'' [PRA \textbf{64}, 062106 (2001)]}


\author{D. Salgado}
\email[]{david.salgado@uam.es}
\affiliation{Dpto. F\'{\i}sica Te\'{o}rica, Universidad Aut\'{o}noma de Madrid\\
28049 Cantoblanco, Madrid (Spain)}

\author{J.L. S\'{a}nchez-G\'{o}mez}
\email[]{jl.sanchezgomez@uam.es}
\altaffiliation{Permanent Address}
\affiliation{Dpto. F\'{\i}sica Te\'{o}rica, Universidad Aut\'{o}noma de Madrid\\
28049 Cantoblanco, Madrid (Spain)}


\date{\today}

\begin{abstract}
We correct a mistake in a result reported in \cite{SteBuz01a}, where it is rightfully argued that initial correlations between a system and its environment may render the system reduced dynamics not completely positive. We prove how not only these initial correlations but also the specific joint dynamics does play a significant role in the question of the complete positivy of the reduced dynamics.
\end{abstract}

\pacs{03.65.Yz}
\keywords{Reduced Dynamics, Complete Positivity, Initial Correlations}

\maketitle


Recently, in the work \cite{SteBuz01a}, \v{S}telmachovi\v{c} and Bu\v{z}ek rightfully argue that in the case of initial quantum correlations between a physical system and its environment, the reduced dynamics of the system might not be expressed in the Kraus form, hence it might not be completely positive. They provide an elementary example of two qubits subjected to a controlled-NOT Hamiltonian evolution and show that in the cases of different initial conditions, namely correlated and uncorrelated conditions, the reduced density operators for one of the qubits are different in both cases. The purpose of this Comment is to prove that the example, as it appears in their work, is wrong, though, we show below, it can still be used to support their correct analysis and to support the statement that not only the initial correlations but also the particular choice of joint dynamics play a significant role in the Kraus representation of the reduced dynamics.\\Starting with the C-NOT Hamiltonian $H=\sigma_{x}\otimes\frac{1}{2}\left(\mathbb{I}-\sigma_{z}\right)+\mathbb{I}\otimes\frac{1}{2}\left(\mathbb{I}+\sigma_{z}\right)$, they find the reduced density matrix $\rho_{A}$ of qubit $A$ at time $t=\frac{\pi}{2}$ with two different initial conditions:

\begin{subequations}
\begin{eqnarray}
\rho_{AB}^{(1)}&=&|\alpha|^{2}|00\rangle\langle00|+|\beta|^{2}|11\rangle\langle11|\\
\rho_{AB}^{(2)}&=&\left(\alpha|00\rangle+\beta|11\rangle\right)\left(\alpha^{*}\langle00|+\beta^{*}\langle11|\right)
\end{eqnarray}  
\end{subequations}

They claim that at time $t=\frac{\pi}{2}$, despite being initially the same, the reduced density matrix of qubit $A$ in each case is:

\begin{subequations}
\begin{eqnarray}
\label{NonCorr}\rho_{A}^{(1)}(t=\pi/2)&=&\frac{1}{2}\left(\mathbb{I}+\sigma_{z}\right)\\
\label{Corr}\rho_{A}^{(2)}(t=\pi/2)&=&\frac{1}{2}\left[\mathbb{I}+\left(|\alpha|^{2}-|\beta|^{2}\right)\sigma_{z}\right]
\end{eqnarray}
\end{subequations}

Elementarily one can argue that when $\alpha=0$ (then $\beta=1$), the reduced density matrix $\rho_{A}$ must be the same since the joint initial states are the same. But this not so (see eqs. \eqref{NonCorr} and \eqref{Corr}). One can also rapidly show that the density matrices at arbitrary times in each case are given by

\begin{subequations}
\begin{eqnarray}
\rho_{A}^{(1)}(t)&=&|\alpha|^{2}\cos t|00\rangle\langle 00|+\left(|\beta|^{2}+|\alpha|^{2}\sin^{2}t\right)|11\rangle\langle11|+\nonumber\\
&+&i|\alpha|^{2}\cos t\sin t|01\rangle\langle01|+h.c.\\
\rho_{A}^{(2)}(t)&=&|\alpha|^{2}\cos t|00\rangle\langle 00|+\left(|\beta|^{2}+|\alpha|^{2}\sin^{2}t\right)|11\rangle\langle11|+\nonumber\\
&+&\left(i|\alpha|^{2}\cos t\sin t+\alpha\beta^{*}e^{it}\cos t\right)|01\rangle\langle01|+h.c.\nonumber\\
\end{eqnarray}
\end{subequations}

\noindent which clearly show that the original claim was wrong. In particular it is remarkable that the diagonal terms are the same, they do not depend on the initial correlations. This means that to detect the effect of initial correlations one should perform an adequate interference experiment (a measurement in a basis other than the computational one in the given example). Here we underline the relevance of the joint system-environment dynamics to obtain a reduced density matrix which cannot be expressed in Kraus form (thus the reduced dynamics not being completely positive) and give a sufficient condition to have a completely positive reduced dynamics even in the case of initial correlations between the system and the environment.\\
In \cite{Pec94a} it was discussed that when there exist initial correlations between a system and its environment, the system reduced dynamics need not be completely positive, a result which can be clearly seen in \cite{SteBuz01a} in general terms. We argue that the question of complete positivity does not only depend on the initial correlations but also on the specific kind of joint dynamics. Indeed we prove that when the joint dynamics is factorisable, then the reduced dynamics is always completely positive, even in the presence of initial correlations between the system and the environment. To prove that consider the extra term coming from initial correlations in the expression of the reduced system density matrix at an arbitrary time $t$ $\delta\rho_{A}(t)\equiv\sum_{\mu}\langle\mu|U_{AB}(t)\gamma'_{ij}\sigma_{i}\otimes\tau_{j}U_{AB}^{\dagger}(t)|\mu\rangle$ (eq. (2.5) in \cite{SteBuz01a}). We show that when $U_{AB}(t)=U_{A}(t)\otimes U_{B}(t)$, then $\delta\rho_{A}(t)=0$ for all $t$. Denoting the linear projectors $P_{\alpha\beta}\equiv|\alpha\rangle\langle\beta|$, it is straightforward to prove that  $\langle\alpha|\delta\rho_{A}(t)|\beta\rangle=\textrm{Tr}\left[\left(U_{AB}^{\dagger}(t)(P_{\beta\alpha}\otimes\mathbb{I})U_{AB}(t)\right)\gamma'_{ij}\sigma_{i}\otimes\tau_{j}\right]$, then since the generators $\tau_{k}$ are traceless for each $k$ and since $U_{B}(t)$ is unitary for each $t$,

\begin{widetext}
\begin{equation}
\textrm{Tr}\left[\left(U_{AB}^{\dagger}(t)(P_{\beta\alpha}\otimes\mathbb{I})U_{AB}(t)\right)\gamma'_{ij}\sigma_{i}\otimes\tau_{j}\right]=\gamma'_{ij}\textrm{Tr}\left(U_{A}^{\dagger}(t)P_{\beta\alpha}U_{A}(t)\sigma_{i}\right)\textrm{Tr}\left(\tau_{j}\right)=0
\end{equation}
\end{widetext}

This result clearly shows that both initial correlations (as it was already known) and the joint dynamics play a relevant role in the question of the complete positivity of the reduced dynamics. In fact, the latter can even be more determinant in some circunstances as we have shown. In the example provided in \cite{SteBuz01a} it can be straightforwardly seen that the dynamics is not factorisable and thus the preceding result cannot be applied. To find necessary conditions to obtain a completely positive reduced dynamics independently of initial correlations stays as an open problem.

\end{document}